\documentclass[twocolumn,superscriptaddress,amsmath,amssymb,aps,longbibliography,nofootinbib,prx]{revtex4-2}
\usepackage[colorlinks=true,urlcolor=blue,citecolor=blue,linkcolor=blue]{hyperref}
\usepackage{txfonts}
\usepackage{graphicx}
\usepackage{bm}
\usepackage{booktabs}

\usepackage{braket}
\usepackage{dcolumn} 

\renewcommand{\vec}[1]{\bm{#1}}

\newcommand{\Eq}[1]{Eq.~(\ref{#1})}
\newcommand{\Fig}[1]{Fig.~\ref{#1}}
\newcommand{\Tr}{\mathrm{Tr}}

\begin{document}

\title{Ab-initio study of interacting fermions at finite temperature with neural canonical transformation}

\author{Hao Xie}
\affiliation{Institute of Physics, Chinese Academy of Sciences, Beijing 100190, China}
\affiliation{University of Chinese Academy of Sciences, Beijing 100049, China}

\author{Linfeng Zhang}
\email{linfeng.zhang.zlf@gmail.com}
\affiliation{AI for Science Institute, Beijing 100080, China}
\affiliation{DP Technology, Beijing 100080, China}

\author{Lei Wang}
\email{wanglei@iphy.ac.cn}
\affiliation{Institute of Physics, Chinese Academy of Sciences, Beijing 100190, China}
\affiliation{Songshan Lake Materials Laboratory, Dongguan, Guangdong 523808, China}

\date{\today}

\begin{abstract}
    We present a variational density matrix approach to the thermal properties of interacting fermions in the continuum. The variational density matrix is parametrized by a permutation equivariant many-body unitary transformation together with a discrete probabilistic model. The unitary transformation is implemented as a quantum counterpart of neural canonical transformation, which incorporates correlation effects via a flow of fermion coordinates. As the first application, we study electrons in a two-dimensional quantum dot with an interaction-induced crossover from Fermi liquid to Wigner molecule. The present approach provides accurate results in the low-temperature regime, where conventional quantum Monte Carlo methods face severe difficulties due to the fermion sign problem. The approach is general and flexible for further extensions, thus holds the promise to deliver new physical results on strongly correlated fermions in the context of ultracold quantum gases, condensed matter, and warm dense matter physics.
\end{abstract}

\maketitle

\section{Introduction}
Consider an interacting quantum system of $N$ fermions in a $d$-dimensional continuous space with the generic Hamiltonian
\begin{equation}
    H = -\frac{1}{2} \nabla^2 + V(\vec{x}),
    \label{eq: Hamiltonian}
\end{equation}
where $\nabla^2  = \sum_{i=1}^N  \nabla_i^2$ and $V(\vec{x}) = \sum_{i=1}^N v^{(1)}(\vec{r}_i) + \sum_{i<j}^N v^{(2)}(\vec{r}_i - \vec{r}_j)$ consists of one- and two-body potentials. We use the short-hand notation $\vec{x} \equiv (\vec{r}_1, \vec{r}_2,\ldots, \vec{r}_N) \in \mathbb{R}^{dN}$ to collectively denote all fermion coordinates. We also assume appropriate natural units so that constants like fermion mass and the Planck constant can be omitted. We would like to study thermodynamic properties of the system encoded in the partition function $Z = \Tr e^{- \beta H}$ at inverse temperature $\beta=1/k_B T$, which are relevant to a broad range of problems including ultracold Fermi gases~\cite{RevModPhys.80.1215}, condensed matter~\cite{RevModPhys.74.1283}, and warm dense matter~\cite{DORNHEIM20181}.
  
Unfortunately, accurate ab-initio study of \Eq{eq: Hamiltonian} at finite temperatures is generally difficult. As a typical workhorse, quantum Monte Carlo methods suffer from the notorious fermion sign problem at low temperatures~\cite{PhysRevB.41.9301, cepeiley1996path, PhysRevE.100.023307,Lee2021a}. There are extensions of ground state quantum chemistry methods to finite-temperature, e.g.~\cite{Hermes2015}. On the other hand, a fundamental principle to solve quantum system at finite temperature is to minimize the variational free energy
\begin{equation}
    F = \frac{1}{\beta} \Tr(\rho \ln \rho) + \Tr(\rho H)
    \label{eq: variational free energy}
\end{equation}
with respect to a variational density matrix $\rho$.  
The two terms in \Eq{eq: variational free energy} correspond to the entropy and energy of the system, respectively. It can be shown that $F \geqslant -\frac{1}{\beta} \ln Z$ is a variational upper bound of the true free energy, where the equality holds only when $\rho$ coincides with the exact density matrix $\frac{1}{Z} e^{-\beta H}$ of the system~\cite{Huber1968}. 

There are a number of physical constraints on the variational density matrix $\rho$. Besides basic properties like Hermitian ($\rho^\dagger=\rho$), positive definiteness ($\rho \succ 0$) and normalization ($\Tr{\rho}=1$), it should also be antisymmetric $\langle \mathcal{P} \vec{x} | \rho | \vec{x}^{\prime} \rangle = (-1)^{\mathcal{P}} \langle \vec{x} | \rho | \vec{x}^{\prime} \rangle$ with respect to permutations $\mathcal{P}\vec{x} \equiv ( \vec{r}_{\mathcal{P}1}, \vec{r}_{\mathcal{P}2},\ldots, \vec{r}_{\mathcal{P}N} )$ of the fermion coordinates. The challenge is then to devise a tractable computational scheme to perform optimization of $\rho$ within such a constrained space. In practice, the entropy term in \Eq{eq: variational free energy} often turns out to be difficult to compute~\cite{PhysRevE.61.3470,PhysRevLett.117.115701}. As a result, most of the previous variational density matrix studies resort to an alternative imaginary-time evolution approach~\cite{PhysRevE.61.3470, doi:10.1063/1.3592777, doi:10.7566/JPSJ.85.034601, PhysRevB.95.205109}, which is more appropriate at high temperatures.

\begin{figure}[t]
    \includegraphics[width=\columnwidth]{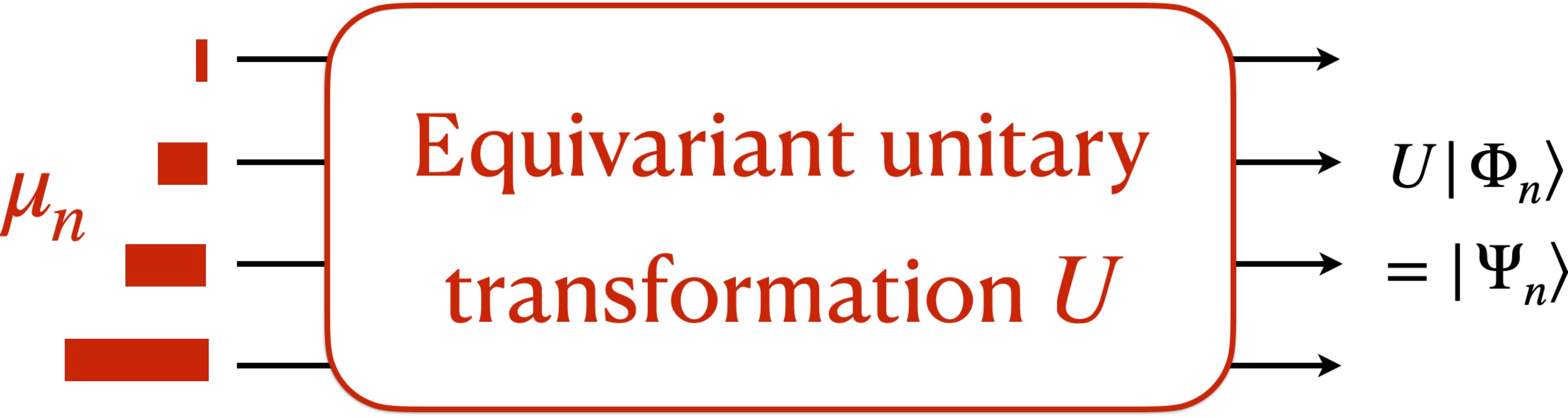}    
    \caption{Architecture of the variational density matrix representation (\ref{eq: density matrix ansatz}) of the present approach. A discrete probabilistic model $\mu_n$ parametrizes the Boltzmann distribution of a many-body basis $|\Psi_n\rangle$. We construct $|\Psi_n\rangle$ by applying a parametrized unitary transformation to a reference basis state $|\Phi_n\rangle$. The unitary transformation corresponds to a permutation equivariant many-body coordinate transformation implemented as a flow of fermion coordinates.  
    }
    \label{fig: schematic_diagram}
\end{figure}

In the low-temperature regime, the variational density matrix can be reasonably represented by a truncated~\footnote{Truncation is necessary for systems in the continuum with infinite-dimensional Hilbert space. Given $N$ fermions and $M$ available single-particle orbitals, the summation in \Eq{eq: density matrix ansatz} involves $M \choose N$ terms, which is exponentially large. Nevertheless, such summation appearing in relevant physical quantities can be estimated via Monta Carlo sampling and will not cause big troubles in practice. See also discussions in Sec.~\ref{sec: application}.} set of low-energy many-body basis states $|\Psi_n\rangle$:
\begin{equation}
    \rho = \sum_{n} \mu_n(\vec{\phi}) |\Psi_n(\vec{\theta})\rangle\langle\Psi_n(\vec{\theta})|,
    \label{eq: density matrix ansatz}
\end{equation}
where $\vec{\phi}$ and $\vec{\theta}$ are variational parameters. A discrete probabilistic model $\mu_n$, which satisfies $0 < \mu_n < 1$ and $\sum_n \mu_n = 1$, is used to parametrize the Boltzmann distribution of the basis states $|\Psi_n\rangle$. On the other hand, we choose to construct $|\Psi_n\rangle$ by applying a unitary transformation to a set of reference basis states $|\Phi_n\rangle$, e.g., the non-interacting Slater determinants: $|\Psi_n(\vec{\theta}) \rangle = U (\vec{\theta}) |\Phi_n\rangle$. To make such an ansatz powerful enough, the unitary transformation should have \textit{many-body} nature, 
so that particle correlations can be effectively introduced into the reference state. 
In addition, it should also preserve permutation antisymmetry of the reference wavefunction, which we will refer to as the \textit{equivariance property}. Overall, the modeling of the variational density matrix in the present approach is illustrated in Figure~\ref{fig: schematic_diagram}.

Parametrizing and optimizing a rich family of equivariant many-body unitary transformations $U$ turn out to be a fairly nontrivial task.
In this paper, we present an elegant solution to this problem by constructing $U$ as unitary representation of the canonical transformation of phase space variables in classical mechanics, extending the previous work on neural canonical transformations~\cite{PhysRevX.10.021020} from classical to quantum domain. The resulting approach naturally generalizes the ground-state variational Monte Carlo (VMC) method~\cite{PhysRev.138.A442, PhysRevB.16.3081} to finite temperatures and is not hindered by the fermion sign problem. Moreover, based on Born's probabilistic interpretation of wavefunctions, the equivariant unitary transformation turns out to be intimately related to equivariant \textit{normalizing flow}~\cite{papamakarios2019normalizing, 9089305}, an important class of generative model developed well within the deep-learning community. This way, one can leverage the latest technical advances in probabilistic modeling to efficiently tackle the thermodynamics of strongly correlated fermions in a fully ab-initio way. 




It is worth mentioning a variety of related works to put the present contribution into a broader perspective. First, there have been various wavefunction ansatzes for \textit{ground-state} VMC calculation of fermions, from the traditional Slater-Jastrow~\cite{PhysRev.98.1479}, backflow~\cite{PhysRev.102.1189, PhysRevLett.47.807, PhysRevB.48.12037, PhysRevE.74.066701} to more recent attempts based on neural networks~\cite{PhysRevLett.122.226401, HAN2019108929, PhysRevResearch.2.033429, Hermann2020, Choo2020, PhysRevB.102.205122, 2020arXiv201107125S, 2021arXiv210312570W, PhysRevLett.127.022502}. 
However, unitary transformations are not considered in these ansatzes, since only a single wavefunction, instead of a whole basis, is needed in this situation. Second, there have been quantum algorithms for thermal properties of model Hamiltonians~\cite{PhysRevA.100.032107, verdon2019quantum,Liu_2021}, which rely on quantum circuits to construct the unitary transformation. However, they still demand advances in quantum technologies to be practically useful. Third, variational free energy studies of statistical mechanics and field theory problems~\cite{PhysRevLett.121.260601,PhysRevLett.122.080602,zhang2018mongeampere, Noeeaaw1147,PhysRevX.10.021020, PhysRevD.100.034515,PhysRevLett.125.121601} can be regarded as the classical counterparts of the present approach. Last but not least, the so-called quantum flow approach~\cite{cranmer2019inferring} also performs a learnable unitary transformation to a single-particle basis. In the many-particle settings considered here, one has to additionally deal with the permutation antisymmetry by imposing the equivariance property into the coordinate transformation carried out by the normalizing flows~\cite{khler2019equivariant,NEURIPS2020_4db73860, bilo2021equivariant}. In this way, normalizing flows have also precisely addressed the open problem envisioned in \cite{eger1963point}: ``The full use of the (coordinate) transform to compute from first principles requires adequate approximation to the Jacobian and the inverse transformation." 

\section{Theory}

The method of constructing unitary transformations in this work is based on the fact that one can establish a one-to-one correspondence between the group of unitary transformations in quantum mechanics and the group of canonical transformations of phase space variables $(\vec{x}, \vec{p})$ in classical mechanics~\cite{doi:10.1063/1.1665805, doi:10.1137/0125024, mexico, BLASZAK201370}. One can develop some basic understanding of the idea by inspecting the infinitesimal structure of these two groups.  In classical mechanics, one can use an arbitrary generating function $G(\vec{x}, \vec{p})$ to define a continuous family 
of canonical transformations via the symplectic evolution $\frac{d \vec{x}}{d\lambda} = \frac{\partial G}{\partial \vec{p}} , \frac{d\vec{p}}{d \lambda}=  - \frac{\partial G}{\partial \vec{x}}$, where $\lambda$ denotes a continuous parameter. When the canonical transformation is quantized, the generating function is converted to a Hermitian operator $\hat{G}$~\footnote{Note that we have put a hat $\hat{\;}$ on operators somewhere in this section to avoid possible confusions.}, and the corresponding unitary transformation then takes the form $U_\lambda = e^{-i\hat{G}\lambda}$. See Appendix \ref{appendix: unitary representation} for more details.

\begin{figure}[t]
    \includegraphics[width=\columnwidth]{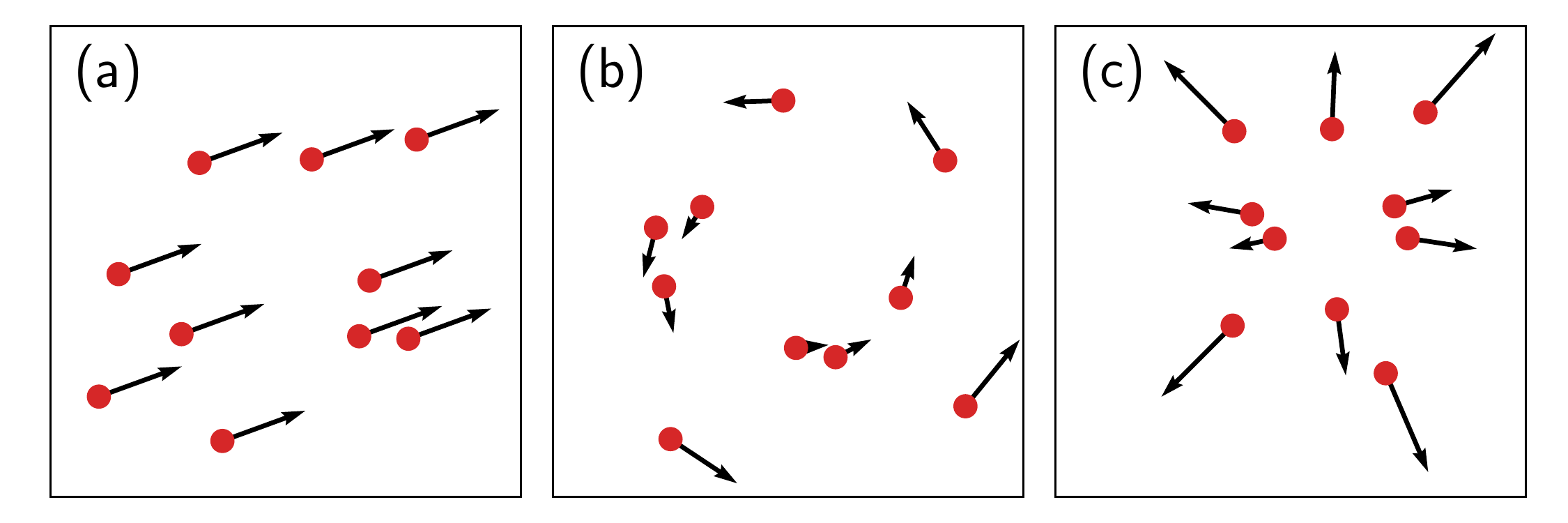}
    \caption{Schematic plot of three different vector fields $\vec{u}$ experienced by the particles. (a) Spatial translation. (b) Spatial rotation. (c) A ``many-body" vector field involving interdependence among the particles.  Evolution under such a vector field will introduce correlation effects.}
    \label{fig: vector_fields}
\end{figure}

An important class of canonical transformation is the so-called \textit{point transformations}, in which the new generalized coordinates depend solely on the old coordinates, not on the old momenta. The generating function of point transformation reads $G = \vec{u}(\vec{x}) \cdot \vec{p}$, where $\vec{u}: (\vec{r}_1, \cdots, \vec{r}_N) \mapsto (\vec{u}_1, \cdots, \vec{u}_N)$ is a function in the $dN$-dimensional coordinate space. The equation of motion followed by the transformed coordinates then takes the form
\begin{equation}
    \frac{d \vec{x}}{d \lambda} = \vec{u}(\vec{x}).
    \label{eq: continuous point transformation} 
\end{equation}
$\vec{u}$ can be intuitively viewed as a vector field that guides all the particles to continuously flow in the coordinate space as the parameter $\lambda$ increases~\footnote{Most generally, the vector field $\vec{u}(\vec{x}, \lambda)$ can also depend on the continuous parameter $\lambda$. The resulting symplectic evolution followed by the particles is then ``time-inhomogeneous''.}. To see this, one can consider the examples of spatial translation and rotation, which can be generated by the total momentum $G = \vec{e} \cdot \sum_{i=1}^N \vec{p}_i$ or angular momentum $G = \vec{n} \cdot \sum_{i=1}^N (\vec{r}_i \times \vec{p}_i)$ along certain directions. The corresponding vector fields are $\vec{u}_i = \vec{e}$ and $\vec{u}_i = \vec{n} \times \vec{r}_i$ respectively, as illustrated in the left and center panel of Figure \ref{fig: vector_fields}. For systems with such spatial symmetries, these transformations would leave the Hamiltonian unaltered. We also note that the vector fields associated with these two examples are ``separable", i.e., the vector field experienced by one particle is independent of the positions of any other particles, thus completely ignores the interactions among them. For practical variational calculations, we need to seek for vector fields $\vec{u}$ that can effectively introduce particle correlations, as illustrated in the right panel of \Fig{fig: vector_fields}.

To study the unitary transformation induced by a general point transformation, we simply employ an anticommutator to make the quantized generator Hermitian~\cite{PhysRev.85.653}: $\hat{G}=\frac{1}{2} \{\vec{u}(\hat{\vec{x}}), \hat{\vec{p}}\}$. Given a set of basis states $\Phi_n(\vec{x})$, the transformed basis wavefunction $\Psi_n(\vec{x}, \lambda) \equiv (U_\lambda \Phi_n)(\vec{x})$ then reads
\begin{equation}
    \Psi_n(\vec{x}, \lambda) = 
    \langle \vec{x} | e^{-\frac{i}{2}\left \{ \vec{u}(\hat{\vec{x}}), \hat{\vec{p}} \right\} \lambda} | \Phi_n \rangle.
    \label{eq: basis wavefunction ansatz}
\end{equation}
To obtain a physical interpretation, it is instructive to differentiate both sides of \Eq{eq: basis wavefunction ansatz} with respect to $\lambda$ to get
\begin{equation}
    \frac{\partial}{\partial \lambda} |\Psi_n(\vec{x}, \lambda)|^2 + \nabla \cdot \left( |\Psi_n(\vec{x}, \lambda)|^2 \vec{u}(\vec{x}) \right) = 0,
    \label{eq: continuity equation} 
\end{equation}
which has the form of a continuity equation of probability density. See Appendix \ref{appendix: unitary representation} for the derivation details. Pictorially, starting from a family of orthonormal reference states such as Slater determinants, the probability mass of each many-body wavefunction will undergo a continuous evolution guided by the many-body vector field $\vec{u}$. During this process, the particles constantly repel or attract each other and correlation effects are gradually cumulated. Crucially, these states remain orthonormal thanks to the unitary nature of the transformation. We thus obtain a correlated many-body basis in the end of the evolution, which can be used to build up the variational density matrix ansatz \Eq{eq: density matrix ansatz}. Note the particles follow deterministic advection in \Eq{eq: continuity equation} rather than random diffusion. In practice, we also integrate the differential equation for a finite amount of time instead of seeking for steady state solutions as diffusion-based approaches~\cite{doi:10.1063/1.431514,Tabak2010, pmlr-v107-barr20a}.  

Transforming probability density continuously in the coordinate space is precisely the idea of continuous normalizing flow~\cite{Tabak2010, Weinan2017, NEURIPS2018_69386f6b, zhang2018mongeampere}. Specifically, the probability $p_n(\vec{z}) \equiv |\Phi_n(\vec{z})|^2$ associated with the reference state is known as the \textit{base} distribution, while the more complex \textit{model} distribution $q_n(\vec{x}) \equiv |\Psi_n(\vec{x})|^2$ is obtained from the base by applying a learnable diffeomorphism $\vec{f}: \vec{z} \mapsto \vec{x}$ in the $dN$-dimensional coordinate space. Note we have omitted $\lambda$ in the notation to avoid cluttering. In the present setting, one builds the diffeomorphism $\vec{f}$ via the ordinary differential equation (ODE) (\ref{eq: continuous point transformation}). By making use of the change-of-variable formula, the model probability $q_n(\vec{x})$ can be written as
\begin{equation}
    q_n(\vec{x}) = p_n(\vec{f}^{-1}(\vec{x}))
    \left| \det \left( \frac{\partial \vec{f}^{-1}(\vec{x})}{\partial \vec{x}} \right) \right|.
\end{equation}
Taking the square root of both sides yields a more explicit expression for the basis wavefunction~\footnote{Such a sloppy ``derivation" is fairly intuitive, yet not satisfying enough for mathematical rigor. In particular, \Eq{eq: basis wavefunction ansatz explicit form} implicitly assumes the wavefunction ansatz $\Psi_n(\vec{x})$ has exactly the same \emph{phase} as $\Phi_n(\vec{z})$, up to a spatial deformation brought by the transformation $\vec{z} \rightarrow \vec{x}$. Fortunately, this is indeed the case. One easy way to see this is from \Eq{eq: basis wavefunction ansatz}: note $\hat{\vec{p}} = -i \nabla$, thus the exponential operator acting on the base wavefunction $\Phi_n$ is actually \emph{real valued}.}:
\begin{equation}
    \Psi_n(\vec{x}) = \Phi_n(\vec{f}^{-1}(\vec{x}))
    \left| \det \left( \frac{\partial \vec{f}^{-1}(\vec{x})}{\partial \vec{x}} \right) \right|^\frac{1}{2}.
    \label{eq: basis wavefunction ansatz explicit form}
\end{equation}
Albeit not so evident at the first sight, Eqs. (\ref{eq: basis wavefunction ansatz explicit form}) and (\ref{eq: basis wavefunction ansatz}) are completely equivalent, which can be rigorously proved by more formally establishing the unitary representation of point transformations; see Appendix \ref{appendix: unitary representation}. In practice, the diffeomorphism $\vec{f}$ can be constructed by composing a sequence of point transformations, which is similar to the iterative backflow approach in ground state variational calculations~\cite{PhysRevB.91.115106, PhysRevLett.120.205302}. However, an important difference of \Eq{eq: basis wavefunction ansatz explicit form} from the ground-state backflow wavefunction ansatzes is the presence of a Jacobian determinant factor. This factor is crucial to guarantee orthonormality of the basis states, which is an essential ingredient for the present finite-temperature approach.

Finally, as a many-fermion wavefunction, $\Psi_n(\vec{x})$ should satisfy the permutation antisymmetry property. Since this property holds already for the base $\Phi_n(\vec{z})$, the only requirement is the unitary transformation appearing in \Eq{eq: basis wavefunction ansatz} being \textit{permutation equivariant}: that is, it should commute with the particle permutation operator. This can be achieved simply by requiring the many-body vector field $\vec{u}$ to be equivariant too, which means that the permutation of particle positions will result in the same permutation of the vector fields they experience:
\begin{equation}
    \vec{u}(\mathcal{P} \vec{x}) = \mathcal{P} \vec{u}(\vec{x}).
    \label{eq: equivariant vector field}
\end{equation}
Intuitively, the indistinguishability of the particles is maintained throughout the continuous flow in the coordinate space, since one cannot label them by using the vector fields they experience at any time. The probability density $q_n(\vec{x}) = |\Psi_n(\vec{x})|^2$ associated with the transformed wavefunction, on the other hand, is \textit{invariant} under particle permutations~\cite{khler2019equivariant,NEURIPS2020_4db73860, bilo2021equivariant}. 
A notable feature is that $q_n(\vec{x})$ inherits nodal lines from the fermionic reference state $p_n(\vec{z}) = |\Phi_n(\vec{z})|^2$ and, crucially, these nodal lines are deformed by the flow transformation.  

To parametrize the permutation equivariant vector field $\vec{u}$, one can leverage many recent advances in natural language processing~\cite{NIPS2017_3f5ee243}, molecular simulation~\cite{NEURIPS2018_e2ad76f2,khler2019equivariant,doi:10.1063/5.0018903}, and point set modeling~\cite{NEURIPS2020_4db73860, bilo2021equivariant}. Moreover, permutation equivariant functions have also been used in various ground-state VMC calculations~\cite{PhysRevE.74.066701, Hermann2020, PhysRevResearch.2.033429}. Consequently, one can naturally port these efforts into the present framework almost without any modifications: just use the permutation equivariant layer as the vector field $\vec{u}$ to drive the flow. 

\section{Implementation\label{sec: implementation}}

Substitution of the density matrix ansatz (\ref{eq: density matrix ansatz}) into \Eq{eq: variational free energy} yields the following estimator of the variational free energy:
\begin{equation}
    F = \mathop{\mathbb{E}}_{n \sim \mu_n} \left[ \frac{1}{\beta} \ln \mu_n + \mathop{\mathbb{E}}_{\vec{x} \sim q_n(\vec{x})} \left[ E_n^\textrm{loc}(\vec{x}) \right] \right].
    \label{eq: variational free energy parametrized form}
\end{equation}
Notice the entropy term depends solely on the state occupation probability $\mu_n$ and can be easily computed, which is a direct consequence of orthonormality of the basis states (\ref{eq: basis wavefunction ansatz explicit form}). The second term consists of the local energy associated with each basis state:
\begin{align}
    E_n^\textrm{loc}(\vec{x}) &\equiv \frac{H \Psi_n(\vec{x})}{\Psi_n(\vec{x})} \nonumber \\
    &= - \frac{1}{4} \nabla^2 \ln q_n(\vec{x}) 
       - \frac{1}{8} \left(\nabla \ln q_n(\vec{x})\right)^2 + V(\vec{x}).
    \label{eq: local energy}
\end{align}
In \Eq{eq: variational free energy parametrized form} the two-fold expectations correspond to classical thermal average of the Boltzmann distribution and quantum expectation according to the Born rule of wavefunction amplitudes. In the limit $\beta \rightarrow \infty$, only the energy term survives and one naturally restores the ground-state VMC method. 

The gradients of \Eq{eq: variational free energy parametrized form} with respect to the parameters $\vec{\phi}$ and $\vec{\theta}$, which appear in the classical and quantum distributions $\mu_n$ and $q_n(\vec{x})$ respectively, have the following forms:
\begin{subequations}
    \begin{align}
        \nabla_{\vec{\phi}} F &= \mathop{\mathbb{E}}_{n \sim \mu_n} \left[
        \left( \frac{1}{\beta} \ln \mu_n + \mathop{\mathbb{E}}_{\vec{x} \sim q_n(\vec{x})} \left[ E_n^\textrm{loc}(\vec{x}) \right] \right)
        \nabla_{\vec{\phi}} \ln \mu_n \right], \\
        \nabla_{\vec{\theta}} F &= \mathop{\mathbb{E}}_{n \sim \mu_n} \mathop{\mathbb{E}}_{\vec{x} \sim q_n(\vec{x})}
        \left[ E_n^\textrm{loc}(\vec{x}) \nabla_{\vec{\theta}} \ln q_n(\vec{x}) \right].
    \end{align}
    \label{eq: gradients of F}
\end{subequations}
For both estimators we employ the control variate method~\cite{JMLR:v21:19-346,PhysRevLett.122.080602,Liu_2021} to further reduce their variances. An important observation is that only the non-negative probability density $q_n(\vec{x}) = |\Psi_n(\vec{x})|^2$ associated with the wavefunction is involved in the calculation. This is a satisfying feature of working directly in the continuum rather than on a finite basis set or lattice~\cite{Choo2020,Carleo2017}: one can deal with the quantum many-body problem completely within the framework of probabilistic modeling. 
Nevertheless, the sign structure of the fermion wavefunction $\Psi_n(\vec{x})$ is still important and relevant for the calculation of off-diagonal physical observables such as correlation function and momentum distribution. 

In practice, Eqs.~(\ref{eq: variational free energy parametrized form}) and (\ref{eq: gradients of F}) are estimated by sampling a batch of pairs $(n, \vec{x})$ from the joint distribution $\mu_n q_n(\vec{x})$ following the ancestral sampling strategy. In particular, to sample coordinates $\vec{x}$, one can start from samples $\vec{z}$ from the prior distribution $p_n(\vec{z})$ (e.g., via Markov chain Monte Carlo) and evolve them according to the ODE (\ref{eq: continuous point transformation}). The log-likelihood $\ln q_n(\vec{x})$ appearing in the local energy~(\ref{eq: local energy}) and gradient estimators~(\ref{eq: gradients of F}) is evaluated by integrating \Eq{eq: continuous point transformation} jointly with the following ODE~\cite{NEURIPS2018_69386f6b,zhang2018mongeampere}: 
\begin{equation}
    \frac{d \ln q_n}{d \lambda} = - \nabla \cdot \vec{u}(\vec{x}).
    \label{eq:dlogqdt}
\end{equation}
To understand this, one can rewrite the continuity equation (\ref{eq: continuity equation}) in the form $\left( \frac{\partial}{\partial \lambda} + \vec{u}(\vec{x}) \cdot \nabla \right) \ln q_n(\vec{x}, \lambda) = -\nabla \cdot \vec{u}(\vec{x})$ and note that $\frac{d}{d \lambda} \equiv \frac{\partial}{\partial \lambda} + \vec{u}(\vec{x}) \cdot \nabla$ is the material derivative associated with the sample $\vec{x}$. Furthermore, the gradient and laplacian operations appearing in (\ref{eq: local energy}) and (\ref{eq: gradients of F}) can be accurately and efficiently computed by differentiating through the ODE integration using automatic differentiation~\cite{JMLR:v18:17-468}, where the adjoint method with constant memory cost turns out to be useful~\cite{NEURIPS2018_69386f6b}. Our code implementation based on PyTorch~\cite{NEURIPS2019_9015} is publicly available~\cite{github}.

\section{Application: electrons in two-dimensional quantum dot\label{sec: application}}

\begin{figure}[t] 
    \includegraphics[width=\columnwidth]{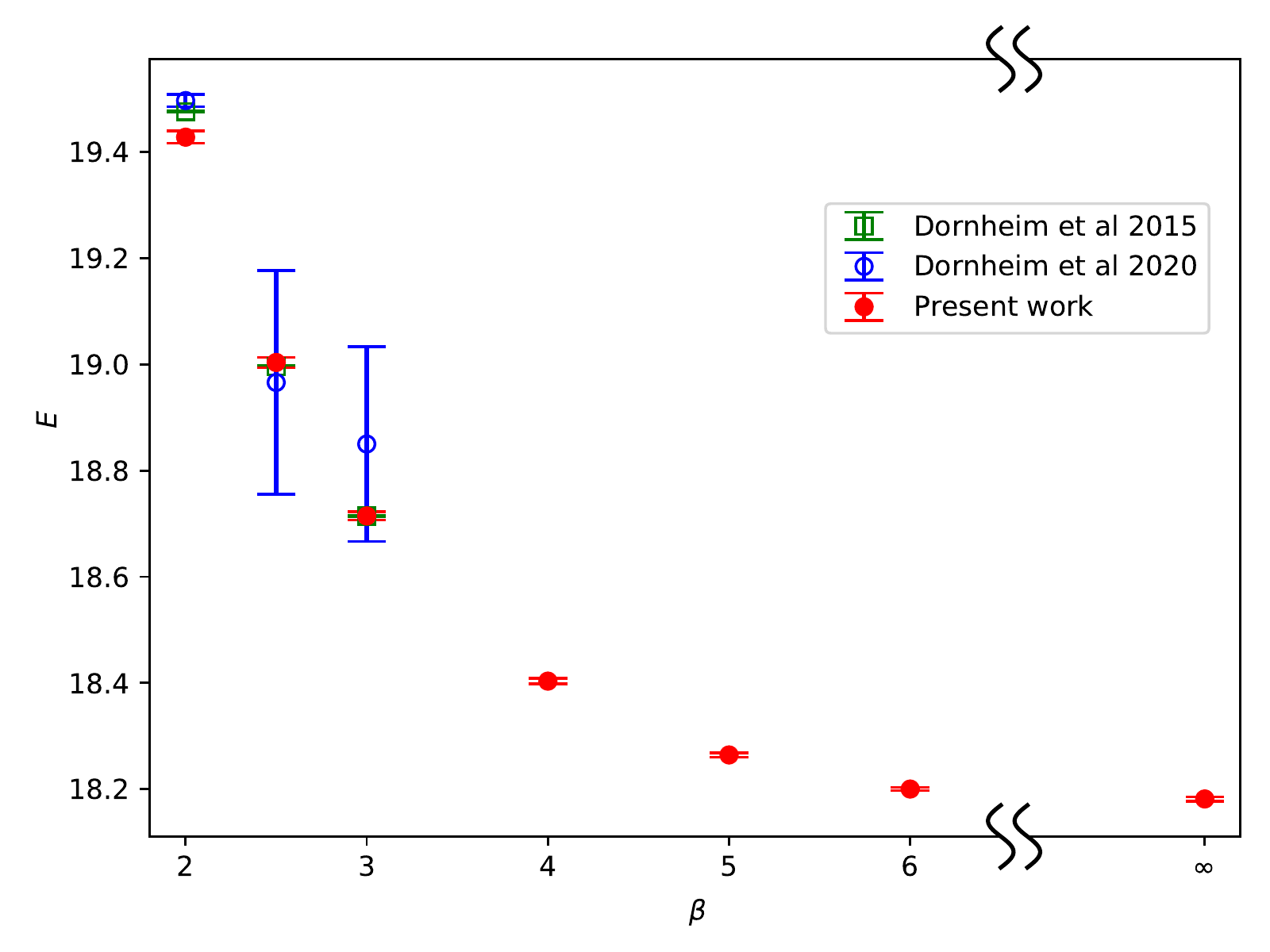}
    \caption{Energy $E$ versus inverse temperature $\beta$ for $6$ spin-polarized electrons in a two-dimensional quantum dot with $\kappa = 0.5$. The green and blue points are benchmark data from two different variants of PIMC~\cite{Dornheim_2015, doi:10.1063/5.0030760}, while the red points are results of the present approach, including the zero-temperature limit.
    }
    \label{fig: energy_data}
\end{figure}

We demonstrate the capability of the present approach by studying electrons in a two-dimensional quantum dot. The one- and two-body potentials take the form of a harmonic trap and repulsive Coulomb interaction, respectively:
\begin{equation}
    v^{(1)}(\vec{r}) = \frac{1}{2} \vec{r}^2, \, v^{(2)}(\vec{r} - \vec{r}^\prime) = \frac{\kappa}{|\vec{r} - \vec{r}^\prime|},
    \label{eq: quantum dots}
\end{equation}
where $\kappa > 0$ is the interaction strength. Despite its simplicity, this model shows rich phenomena due to the interplay of interaction and temperature effects. In particular, as $\kappa$ increases, the Fermi liquid picture based on the concept of quasiparticles would eventually break down. The resulting phase is usually characterized as a \textit{Wigner molecule}~\cite{RevModPhys.74.1283}, where the kinetic motion of electrons is largely frozen, and the spatial density distribution would typically exhibit a shell structure. There have been a large number of numerical studies focusing on its ground-state~\cite{PhysRevB.65.075309, PhysRevB.62.8108, PhysRevB.62.8120} and finite-temperature properties~\cite{PhysRevLett.81.4533, PhysRevLett.82.3320, PhysRevB.62.10207, PhysRevLett.86.3851}. However, there have been no reliable method that works for the entire interaction range at low temperatures. Thus, this problem offers an ideal playground for the present method.

We consider the spin polarized case. The base wavefunctions $\Phi_n(\vec{z})$ are chosen to be Slater determinants of single-electron orbitals obtained simply by eliminating the two-body term $v^{(2)}$. Such Slater determinants constitute an exponentially large set of basis of the many-body Hilbert space. Focusing on low-temperature properties of the system, we carry out a truncation of the basis by including only those within an energy cutoff $E_{\textrm{cut}}$ relative to the noninteracting ground state. In the considered parameter region, we found that $E_{\textrm{cut}} \leqslant 4$ is sufficient to capture most of the finite-temperature effects. Since the corresponding number of basis states is no more than $2000$, we choose to adopt a simple parametrization of the state probabilities $\mu_n(\vec{\phi}) = \frac{e^{\phi_n}}{\sum_m e^{\phi_m}}$ based on the softmax function. Nevertheless, we note that this is not a limiting factor of the present approach because one can capture exponentially large number of basis states by utilizing more sophisticated discrete probabilistic models~\cite{PhysRevLett.122.080602, PhysRevResearch.2.023358, MAL-044,PhysRevX.8.031012}.

The wavefunction ansatz $\Psi_n(\vec{x})$ is generated from the base $\Phi_n(\vec{z})$ by the continuous flow guided by a many-body vector field $\vec{u}$. We take $\vec{u}$ to be of the backflow form~\cite{PhysRev.102.1189, PhysRevLett.47.807, PhysRevB.48.12037, PhysRevE.74.066701} for simplicity and clear physical interpretation: 
\begin{equation}
    \vec{u}_i = \xi(|\vec{r}_i|) \vec{r}_i + \sum_{j \neq i}^N \eta(|\vec{r}_i - \vec{r}_j|) (\vec{r}_i - \vec{r}_j).
    \label{eq: backflow}
\end{equation}
The many-body nature and permutation equivariance of this vector field can be easily confirmed by inspection. The scalar functions $\xi$ and $\eta$ can be referred to as the one- and two-body ``backflow potential", which capture the ``mean field" and electron correlation effects, respectively. Note for a given distance $r$, $\eta(r) > 0$ stands for a repulsive interaction between two electrons, and vice versa; similarly for $\xi(r)$. We parametrize the potentials by two independent neural networks with single hidden layer. Initially, the backflow potentials are set to zero and $\mu_n$ to Boltzmann distribution of the non-interacting base states $\Phi_n(\vec{z})$. The optimization is performed on a batch of 8000 samples using the Adam stochastic gradient descent algorithm~\cite{DBLP:journals/corr/KingmaB14} for $3000$ iteration steps. 

\begin{figure}[t]
    \includegraphics[width=\columnwidth]{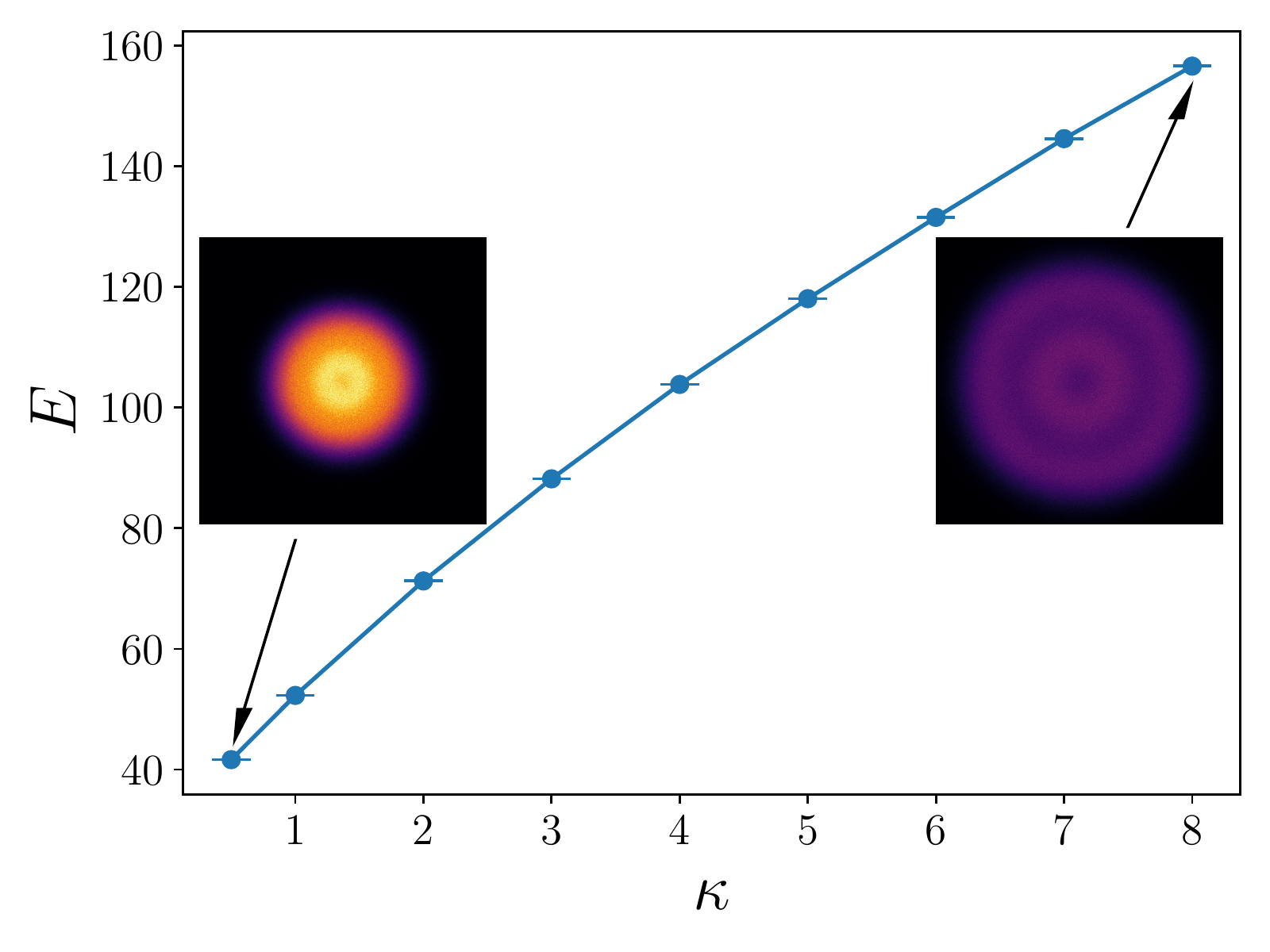}
    \caption{Energy $E$ versus interaction strength $\kappa$ for $10$ spin-polarized electrons in a two-dimensional quantum dot with $\beta=6$. The inset shows the electron density profiles for $\kappa = 0.5$ and $8$.}
    \label{fig: energy_data2}
\end{figure}

As the first benchmark, Figure~\ref{fig: energy_data} shows the temperature dependence of the energy for a system of $N = 6$ spin-polarized electrons with $\kappa = 0.5$. For such a weak interaction, the standard path integral Monte Carlo (PIMC) method is severely hindered by the fermion sign problem, since the electrons are largely delocalized and subject to exchange effects. Consequently, it provides reliable results only at relatively high temperatures $\beta \lesssim 1.5$~\cite{PhysRevE.100.023307}. Variants of PIMC with alleviated fermion sign problem can access slightly lower temperatures~\cite{Dornheim_2015, doi:10.1063/5.0030760}, where our results agree nicely with the benchmark data as shown in the figure. The slight discrepancy at $\beta=2$ is likely due to insufficiently large $E_\textrm{cut}$ in our calculation. On the other hand, the present approach can easily reach even lower temperatures, including the zero-temperature limit $\beta \rightarrow \infty$. We note that alternative Monte Carlo methods based on expansions in the Fock space~\cite{https://doi.org/10.1002/ctpp.201100012,PhysRevB.89.245124} can work more favorably for such weak interactions, but will again suffer from the fermion sign problem in the strong coupling region.

Overall, the present approach serves as a valuable complement of conventional quantum Monte Carlo methods for studying the thermodynamic properties of fermion systems, especially for low temperature, large particle number and intermediate interaction strength~\cite{schoof2017configuration}. To demonstrate this, we perform systematic calculations of $N=10$ spin-polarized electrons at $\beta=6$ for a wide range of $\kappa$ from $0.5$ to $8$. For more benchmark data, see Appendix \ref{appendix: benchmark data}. Figure~\ref{fig: energy_data2} shows the energy dependence on $\kappa$, together with electron density profiles at the two limits $\kappa = 0.5$ and $8$. Notice the density centers around the origin of the trap in the weak coupling regime. On the other hand, stronger repulsive interaction smears out the electron cloud and induces a shell structure, indicating the emergence of the Wigner molecule phase. The observed spatial configuration consisting of two shells for the present parameter settings also agrees with the analysis in the classical limit $\kappa \rightarrow \infty$~\cite{RevModPhys.74.1283}, where quantum fluctuations arising from the kinetic term in \Eq{eq: Hamiltonian} are ignored. Reaching this result in the strong interaction regime where the density profile is qualitatively different from the weak coupling case is a stringent test to the present method.

\begin{figure}[t]
    \includegraphics[width=\columnwidth]{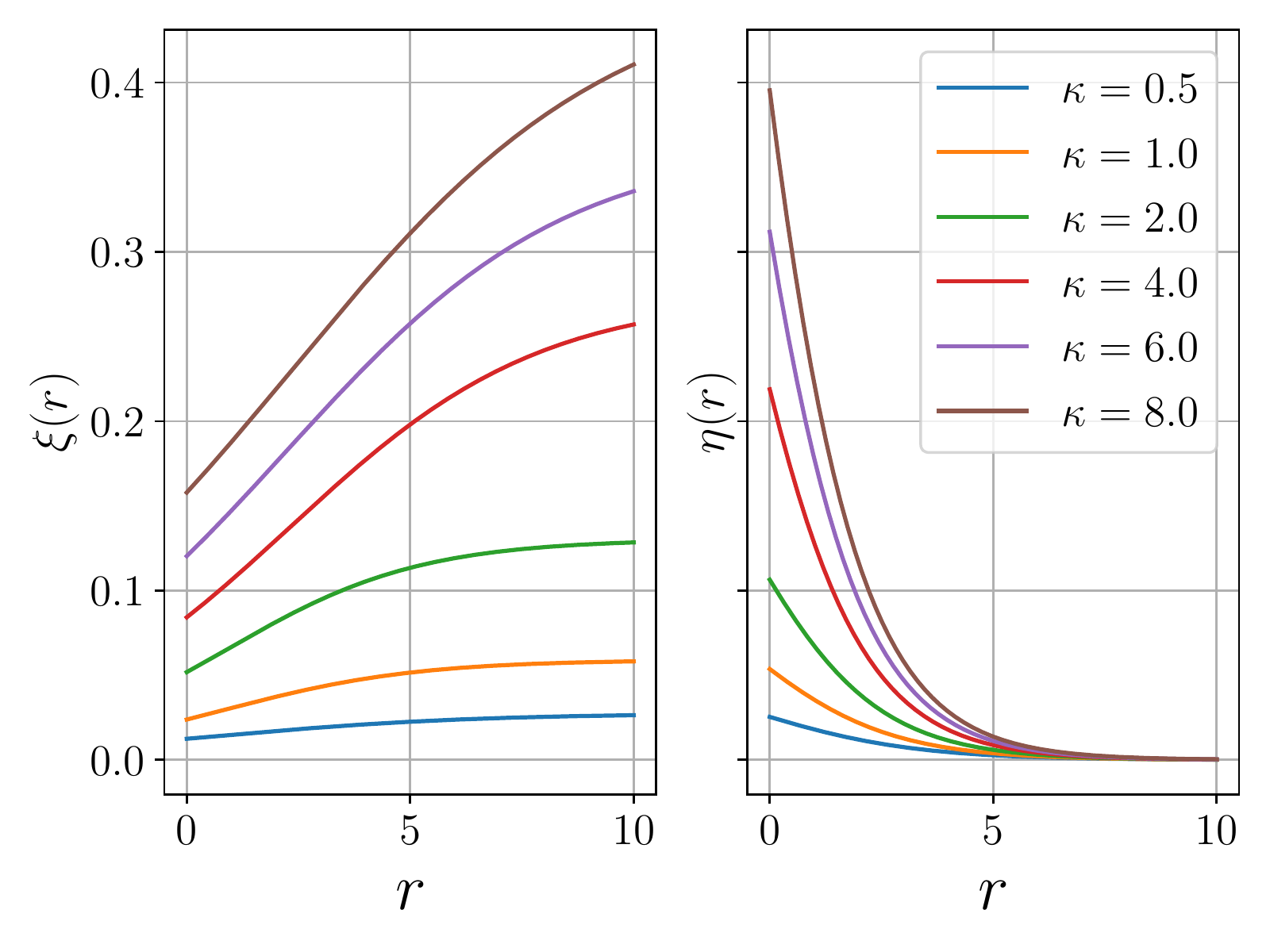}
    \caption{The optimized one- and two-body backflow potentials $\xi$ and $\eta$, respectively, as functions of the distance $r$, for various values of interaction strength $\kappa$ in a quantum dot of $N=10$ spin-polarized electrons with $\beta=6$.}
    \label{fig: backflow}
\end{figure}

\begin{figure*}[t]
    \centering
    \includegraphics[width=\textwidth]{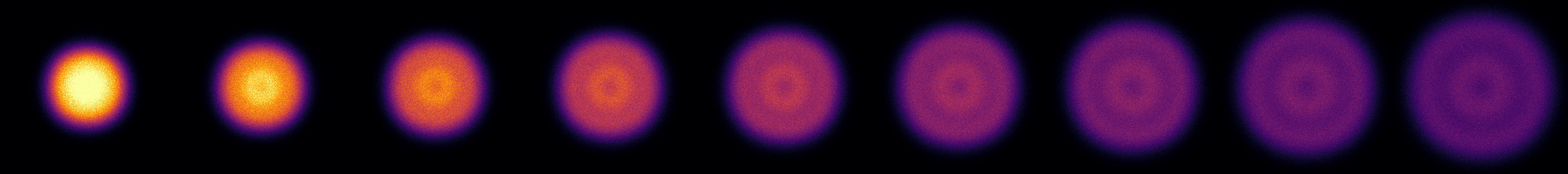}
    \caption{Several equally-spaced snapshots of the electron density along the continuous flow of coordinates at $\lambda=0, 1/8,\ldots, 1$. The system parameters are $N=10, \beta=6$ and $\kappa=8$. Starting from the non-interacting reference state, the electron correlation effects can be gradually introduced, and one can finally reproduce the shell structure characteristic of a Wigner molecule in the strong interaction regime.}
    \label{fig: density2D_snapshots}
\end{figure*}

To obtain the density profiles as shown in \Fig{fig: energy_data2}, one starts from the density of non-interacting reference state consisting of a large number of electron coordinate samples $\vec{z}$, then evolves them according to the continuous flow specified by the ODE (\ref{eq: continuous point transformation}) towards the final spatial distribution of $\vec{x}$, as described previously in Sec.~\ref{sec: implementation}. The initial and final values of the continuous parameter $\lambda$ are conventionally chosen to be $0$ and $1$, respectively, which are treated as fixed hyperparameters of the model. The many-body vector field $\vec{u}$ governing such an evolution process is determined by the backflow potentials $\xi$ and $\eta$, which are shown in Figure \ref{fig: backflow}. Notice the interactions arising from one- and two-body backflow potentials are both repulsive, which can be viewed as the manifestation of electron repulsion at the level of mean field and two-body correlations, respectively. The overall evolution of the electrons is, nevertheless, jointly determined by the two potentials together. Fig.~\ref{fig: backflow} also shows that in the strong interaction regime, the backflow potentials deviate largely from the values of zero in the non-interacting case. To visualize how such strong potentials affect the evolution of electron coordinates, Fig~\ref{fig: density2D_snapshots} shows several density snapshots along the continuous flow at $\lambda=0, 1/8,\ldots, 1$ for $\kappa=8$. The cumulation of electron correlations and onset of the shell structure is clear.

\section{Discussions}

In essence, this work belongs to the large family of canonical transformation approaches for quantum many-body systems, except that the transformation is directly carried out upon particle coordinates instead of a many-body Fock space within the formulation of second quantization~\cite{wegner1994flow, PhysRevD.49.4214, doi:10.1063/1.1508370}. Although the framework of neural canonical transformation is general, we have technically restricted to the subgroup of point transformations to make the calculation tractable, which corresponds to the specific choice $\hat{G}=\frac{1}{2} \{\vec{u}(\hat{\vec{x}}), \hat{\vec{p}}\}$ of the generator. The generalization to arbitrary generators, such as the Hamiltonian $H$ in \Eq{eq: Hamiltonian}, seems fairly nontrivial in practice. In fact, the tractability of such a general calculation implies that it is possible to accurately simulate the real-time evolution $e^{-iHt} |\Phi\rangle$ of any many-body systems. In this perspective, the present approach can also be understood as a short-time variational approximation of adiabatic time evolution towards the thermal equilibrium.

One can take a different view of the limit of the present approach by inspecting the basis wavefunction representation \Eq{eq: basis wavefunction ansatz explicit form}. Compared to ground-state variational ansatzes, the coordinate transformation $\vec{f}$ in this work is implemented by continuous normalizing flow and subject to further limitations due to its invertibility~\cite{NEURIPS2019_21be9a4b, cornish2019relaxing, pmlr-v119-zhang20h, pmlr-v108-kong20a}. Moreover, the continuous flow can only deform the nodal surface of reference states without changing its topology~\cite{PhysRevE.74.066701, PhysRevB.72.075131, PhysRevB.86.115120}. One example is the number of nodal cells of the ground-state wavefunction, which is conjectured to be always two for spatial dimensions $d$ higher than one~\cite{Ceperley1991, PhysRevLett.96.240402, PhysRevB.86.115120, mitas2006fermion}. Similar conjecture has also been proposed for thermal density matrices at low temperatures~\cite{mitas2006fermion}. More thorough characterization of fermion nodes like possible topological obstructions is still lacking and worths further study~\cite{doi:10.1063/1.463296, PhysRevB.78.035104}. In practice, one may remedy these issues by increasing the expressibility of the reference states $\Phi_n(\vec{z})$. For example, one can use more physically plausible reference states than the Slater determinants~\cite{PhysRevLett.96.130201, PhysRevB.77.115112}, or introduce additional parameters into the reference state which are pretrained or trained jointly with the flow transformation.



Although we have made use of continuous normalizing flow in this work, it should be possible to use other class of permutation equivariant normalizing flows~\cite{papamakarios2019normalizing}. Some examples are the partitioned flow~\cite{doi:10.1063/5.0018903} and the invertible residual network~\cite{pmlr-v97-behrmann19a}, which can be more efficient than the present ODE-based implementation. One can also directly carry out Monte Carlo sampling of the electron coordinates other than transforming samples of the reference states. The local energy \Eq{eq: local energy} resembles the score matching loss function~\cite{JMLR:v6:hyvarinen05a} in training generative models, which is known to be expensive to compute. In light of this, advances in efficient score matching training might be beneficial to further reduce the computational efforts when scaling up to larger systems~\cite{pmlr-v115-song20a}. 


Building on these technical improvements in the implementation, a promising future direction is to scale up to larger particle number $N$ and study other correlated fermion systems of fundamental importance, such as the homogeneous electron gas~\cite{doi:10.1063/1.4977920} and dense hydrogen~\cite{PhysRevE.61.3470}. Moreover, similar to what was shown in \cite{Liu_2021}, one can also obtain information about the low-lying excited states of these systems as a byproduct of the thermodynamic calculation.

\begin{acknowledgments}
We thank Yuan Wan, Zi-Long Li, Hao Wu, Zi Cai, Jun Wang, Xiang Chen, and Vincent Moens for useful discussions. We thank Tobias Dornheim for providing the reference data shown in Fig.~\ref{fig: energy_data}. This project is supported by the National Natural Science Foundation of China under Grant No.~11774398, and the Ministry of Science and Technology of China under the Grant No.~2016YFA0300603 and 2016YFA0302400.
\end{acknowledgments}

\appendix


\section{\label{appendix: unitary representation}Unitary representation of canonical transformations}
We elaborate on the one-to-one correspondence between the group of unitary transformations in quantum mechanics and the group of canonical transformations in classical mechanics. Special emphasis will be placed on the subgroup of point transformations, which is the focus of the present work.

\subsection{Basic formulation}
In classical mechanics, a canonical transformation is a smooth bijection from the original set of phase space variables $(\vec{x}, \vec{p})$ to a new one $(\vec{X}(\vec{x}, \vec{p}), \vec{P}(\vec{x}, \vec{p}))$ satisfying the so-called symplectic condition, which is equivalent to saying that all Poisson brackets among the new variables are preserved~\cite{book:2839416}. To study its implication in the realm of quantum mechanics, we should convert the new (as well as old) variables into Hermitian operators $(\vec{X}, \vec{P}) \rightarrow (\hat{\vec{X}}, \hat{\vec{P}})$ following certain quantization procedure. As a result, $(\hat{\vec{X}}, \hat{\vec{P}})$ should satisfy the usual commutation relations of coordinate and momenta as $(\hat{\vec{x}}, \hat{\vec{p}})$. This largely motivates us to reasonably expect the existence of a unitary transformation $U$ that connects the two set of operators together:
\begin{equation}
    \hat{\vec{X}} = U^\dagger \hat{\vec{x}} U, \, \hat{\vec{P}} = U^\dagger \hat{\vec{p}} U.
    \label{eq: operator transformation}
\end{equation}
$U$ can thus be viewed as the unitary representation of the given canonical transformation.

To further clarify the nature of $U$, it is instructive to consider the eigenstate $|\vec{x})$ of the new coordinate operators $\hat{\vec{X}}$ defined as
\begin{equation}
    \hat{\vec{X}} |\vec{x}) = \vec{x} |\vec{x}).
    \label{eq: |x)}
\end{equation}
The essential point is that $|\vec{x})$ constitutes a \textit{different} coordinate basis from the old ones $\hat{\vec{x}} |\vec{x}\rangle = \vec{x} |\vec{x}\rangle$, which is why a slightly different Dirac notation has been used in \Eq{eq: |x)}. By making use of these two basis, the unitary transformation $U$ can then be formally defined as~\cite{doi:10.1063/1.1665805, doi:10.1137/0125024}
\begin{equation}
    U |\vec{x}) = |\vec{x}\rangle.
    \label{eq: U definition formal}
\end{equation}
It is then straightforward to verify the operator transformation relations \Eq{eq: operator transformation}.

The tranformation behavior of $U$ upon the wavefunction $\Psi(\vec{x})$ of any given quantum state reads
\begin{align}
    (U\Psi)(\vec{x}) &= \langle\vec{x}|U|\Psi\rangle \nonumber \\
    &= (\vec{x}|\Psi\rangle = \int d \vec{x}^\prime (\vec{x}|\vec{x}^\prime\rangle \Psi(\vec{x}^\prime).
    \label{eq: wavefunction transformation}
\end{align}
The essential ingredient in evaluating this expression is the state overlap $\langle \vec{x}^\prime | \vec{x} )$, which can be shown to be uniquely determined (up to a phase factor) by the following set of equations~\cite{doi:10.1137/0125024, mexico}:
\begin{subequations}
    \begin{align}
        \hat{\vec{X}}\left( \vec{x}^\prime, -i\frac{\partial}{\partial \vec{x}^\prime} \right) \langle\vec{x}^\prime|\vec{x}) &= \vec{x} \langle\vec{x}^\prime|\vec{x}), \\
        \hat{\vec{P}}\left( \vec{x}^\prime, -i\frac{\partial}{\partial \vec{x}^\prime} \right) \langle\vec{x}^\prime|\vec{x}) &= i \frac{\partial}{\partial \vec{x}} \langle\vec{x}^\prime|\vec{x}).
    \end{align}
    \label{eq: basis overlap equation}
\end{subequations}

Unfortunately, for a general canonical transformation, the computation procedure outlined in Eqs. (\ref{eq: wavefunction transformation}) and (\ref{eq: basis overlap equation}) above can be fairly difficult. We would thus restrict ourselves in the subgroup of point transformations, which can be constructed by the type-2 generating function $F_2(\vec{x}, \vec{P}) = \vec{f}(\vec{x}) \cdot \vec{P}$ via the following implicit relations~\cite{book:2839416}:
\begin{equation}
    \vec{X} = \frac{\partial F_2}{\partial \vec{P}}, \, \vec{p} = \frac{\partial F_2}{\partial \vec{x}}.
    \label{eq: type-2 generating function}
\end{equation}
The resulting transformation formula can be obtained by some simple manipulations:
\begin{equation}
    \vec{X} = \vec{f}(\vec{x}), \, \vec{P} = \left(\frac{\partial \vec{f}}{\partial \vec{x}}\right)^{-T} \vec{p}.
    \label{eq: point transformation}
\end{equation}
In other words, the new coordinates depend only on the old coordinates through a bijective map $\vec{f}$, while the momenta are transformed in a covariant way so as to preserve the Poisson brackets. When such a point transformation is quantized, the corresponding expression for the state overlap $\langle \vec{x}^\prime | \vec{x} )$ turns out to be simple:
\begin{equation}
    \langle\vec{x}^\prime|\vec{x}) = \delta\left(\vec{x}^\prime - \vec{f}^{-1}(\vec{x})\right)
\left| \det \left( \frac{\partial \vec{f}^{-1}(\vec{x})}{\partial \vec{x}} \right) \right|^{\frac{1}{2}}.
    \label{eq: basis overlap for point transformation}
\end{equation}
The correctness of this result can of course be verified by plugging it into \Eq{eq: basis overlap equation}. One can, however, get the intuitive feeling by noting that $\hat{\vec{X}} \equiv \vec{f}(\hat{\vec{x}})$ shares the same set of eigenstates as $\hat{\vec{x}}$, i.e., the Dirac delta function, while the additional Jacobian determinant factor is present to make the integration measure in the completeness relation $\int d \vec{x} |\vec{x})(\vec{x}| = 1$ as expected. By substituting (\ref{eq: basis overlap for point transformation}) into \Eq{eq: wavefunction transformation}, we finally obtain the transformed wavefunction through a point transformation as follows:
\begin{equation}
    (U\Psi)(\vec{x}) = \Psi\left(\vec{f}^{-1}(\vec{x})\right)
\left| \det \left( \frac{\partial \vec{f}^{-1}(\vec{x})}{\partial \vec{x}} \right) \right|^{\frac{1}{2}}.
    \label{eq: wavefunction for point transformation}
\end{equation}
This result is essentially \Eq{eq: basis wavefunction ansatz explicit form} in the main text, where it is intuitively obtained from the perspective of normalizing flow.

\subsection{Infinitesimal unitary transformations}
The characteristics of the unitary representation of canonical transformations formulated above can be more clearly revealed by studying the infinitesimal behavior in the vicinity of identity transformation, which is a common practice in physics and of great theoretical importance. An infinitesimal canonical transformation can be constructed by the type-2 generating function $F_2(\vec{x}, \vec{P}) = \vec{x} \cdot \vec{P} + d \lambda \, G(\vec{x}, \vec{P})$, where the two terms correspond to identity transformation and infinitesimal perturbation, respectively. Substituting this expression into \Eq{eq: type-2 generating function} and retaining only lowest-order contributions, one could then obtain a continuous family $(\vec{x}(\lambda), \vec{p}(\lambda))$ of canonical transformations specified by the Hamilton's equations of motion:
\begin{equation}
    \frac{d \vec{x}}{d \lambda} = \frac{\partial G}{\partial \vec{p}}, \, \frac{d \vec{p}}{d \lambda} = - \frac{\partial G}{\partial \vec{x}}.
    \label{eq: continuous canonical transformation}
\end{equation}
The function $G(\vec{x}, \vec{p})$ is usually also called the generating function. When such a canonical transformation is quantized, \Eq{eq: continuous canonical transformation} is naturally replaced by the Heisenberg equations of motion. This observation is essential: in light of \Eq{eq: operator transformation}, the corresponding unitary transformation can be immediately recognized as
\begin{equation}
    U = e^{-i\hat{G}\lambda},
    \label{eq: U definition explicit}
\end{equation}
where the generator $\hat{G}$ should be ensured to be Hermitian by the quantization procedure.

Equation~(\ref{eq: U definition explicit}) is clearly a more explicit and meaningful characterization of the unitary transformation than the formal definition (\ref{eq: U definition formal}). In particular, the state overlap appearing in \Eq{eq: wavefunction transformation} corresponds precisely to the propagator $( \vec{x} | \vec{x}^\prime \rangle = \langle \vec{x} | e^{-i\hat{G}\lambda} | \vec{x}^\prime \rangle$. Such a propagator is difficult to evaluate in general cases, so we again concentrate only on the point transformations as in the previous section. Within the formulation presented here, a point transformation corresponds to the choice $G(\vec{x}, \vec{p}) = \vec{u}(\vec{x}) \cdot \vec{p}$, which leads to the type-2 generating function
\begin{equation}
    F_2(\vec{x}, \vec{P}) = \left( \vec{x} + d \lambda \, \vec{u}(\vec{x}) \right) \cdot \vec{P}.
    \label{eq: type-2 generating function infinitesimal form}
\end{equation}
To obtain the corresponding unitary representation, we employ a simple operator symmetrization $\hat{G} = \frac{1}{2} \{\vec{u}(\hat{\vec{x}}), \hat{\vec{p}}\}$ to make the generator Hermitian, as mentioned in the main text. Comparing \Eq{eq: type-2 generating function infinitesimal form} with the form $F_2(\vec{x}, \vec{P}) = \vec{f}(\vec{x}) \cdot \vec{P}$ discussed in the previous section, one can readily reach the conclusion that the transformed wavefunction $(U \Psi) (\vec{x}) = \langle \vec{x} | e^{-\frac{i}{2} \{\vec{u}(\hat{\vec{x}}), \hat{\vec{p}}\} \lambda} | \Psi\rangle$ on a quantum state $\Psi(\vec{x})$ can be equivalently written in the form of \Eq{eq: wavefunction for point transformation}, where the coordinate bijection $\vec{f}$ is specified by the ordinary differential equation
\begin{equation}
    \frac{d \vec{x}}{d \lambda} = \vec{u}(\vec{x}).
\end{equation}
We thus rigorously show the equivalence of the two basis wavefunction ansatz expressions Eqs. (\ref{eq: basis wavefunction ansatz}) and (\ref{eq: basis wavefunction ansatz explicit form}), which lies at the core of the finite-temperature approach in this work.

Finally, we give a few guidelines for the derivation of the continuity equation (\ref{eq: continuity equation}) from \Eq{eq: basis wavefunction ansatz} in the main text for readers' convenience. Note that $\hat{\vec{p}} = -i \nabla$, we have
\begin{align}
    \frac{\partial}{\partial \lambda} \Psi_n(\vec{x}, \lambda) &= -\frac{i}{2}\left \{ \vec{u}(\hat{\vec{x}}), \hat{\vec{p}} \right\} \Psi_n(\vec{x}, \lambda) \nonumber \\
    &= -\frac{1}{2} \bigg[ \vec{u}(\vec{x}) \cdot \nabla \Psi_n(\vec{x}, \lambda) + \nabla \cdot \left( \vec{u}(\vec{x}) \Psi_n(\vec{x}, \lambda) \right) \bigg] \nonumber \\
    &= -\vec{u}(\vec{x}) \cdot \nabla \Psi_n(\vec{x}, \lambda) - \frac{1}{2} \Psi_n(\vec{x}, \lambda) \nabla \cdot \vec{u}(\vec{x}).
    \label{eq: continuity equation derivation}
\end{align}
To obtain \Eq{eq: continuity equation}, simply multiply \Eq{eq: continuity equation derivation} by $\Psi_n^\ast(\vec{x}, \lambda)$ and add the resulting equation to its own complex conjugate.

\section{\label{appendix: benchmark data}Some more benchmark data for 2D quantum dot}
\renewcommand{\arraystretch}{1.2} 
\setlength{\tabcolsep}{1em}
\begin{table}[b]
\begin{tabular}{@{}rrrr@{}}\toprule
        $N$ & $\kappa$ &
        \multicolumn{1}{c}{This work} &
        \multicolumn{1}{c}{\textrm{\cite{PhysRevLett.82.3320}}} \\
        \colrule
        3 & 2 & 8.331(3) & 8.37(1) \\
        3 & 4 & 11.070(4) & 11.05(1) \\
        3 & 6 & 13.495(6) & 13.43(1) \\
        3 & 8 & 15.653(7) & 15.59(1) \\
       \cmidrule{1-4}
        4 & 2 & 14.336(4) & 14.30(5) \\
        4 & 4 & 19.517(7) & 19.42(1) \\
        4 & 6 & 24.060(9) & 23.790(12) \\
        4 & 8 & 28.178(12) & 27.823(11) \\
        \cmidrule{1-4}
        6 & 0.5 & 18.179(4) &   -- \\
        6 & 1 & 22.003(6) &  --   \\
        6 & 1.5 & 25.600(8) &  --  \\
        6 & 2 & 28.994(9) &  --  \\
        6 & 3 & 35.241(10) & -- \\
        6 & 4 & 41.012(11) & --  \\
        6 & 5 & 46.385(13) & --  \\
        6 & 6 & 51.448(13) & --  \\
        6 & 7 & 56.270(16) & --  \\
        6 & 8 & 60.837(15) & 60.42(2) \\
        \bottomrule
\end{tabular}
\end{table}
The following table summarizes our results for the energy of a two-dimensional quantum dot at $\beta=10$, for various electron number $N$ and interaction strength $\kappa$. PIMC results from \cite{PhysRevLett.82.3320} are also listed when available. All data correspond to the fully spin-polarized case. Our finite-temperature calculations indicate that the entropy is negligible for a temperature as low as $\beta=10$, so our energy results can be treated as variational. We anticipate these results (as well as those presented in the main text) can be further improved by adopting better model architecture and optimization schemes. We also note the results reported in Figure 4.8 of \cite{schoof2017configuration} for $\beta=10, N=3, \kappa=2$ show that the data in \cite{PhysRevLett.82.3320} may be subject to slight systematic errors.

\bibliography{references}

\end{document}